\documentclass[twoside,final]{pro12}

\usepackage[utf8]{inputenc}
\usepackage{amsmath}
\usepackage{graphicx}
\usepackage{array}
\usepackage{SIunits}
\usepackage{lscape}
\usepackage{multirow}
\usepackage{float}
\usepackage{natbib}
\usepackage{hyperref}

\DeclareGraphicsExtensions{.pdf,.png}

\head{Pearse C. Murphy et al.}{Automatic SRB detection with NenuFAR}	

\begin{document}

\title{Automatic detection of solar radio bursts in NenuFAR observations}	
\author{Pearse C. Murphy\adress{\textsl LESIA, Observatoire de Paris, 
Université PSL, Sorbonne Université, Université Paris Cité, CNRS, 92190 
Meudon, France}$\,\,$, Baptiste Cecconi$^1$, Carine Briand$^1$, \\and 
St\'ephane Aicardi\adress{\textsl DIO, Observatoire de Paris, Université 
PSL, CNRS, 75014 Paris, France
}}

\maketitle

\begin{abstract}
Solar radio bursts are some of the brightest emissions at radio frequencies 
in the solar system. The emission mechanisms that generate these bursts offer 
a remote insight into physical processes in solar coronal plasma, while fine 
spectral features hint at its underlying turbulent nature. During radio noise 
storms many hundreds of solar radio bursts can occur over the course of a few 
hours. Identifying and classifying solar radio bursts is often done manually 
although a number of automatic algorithms have been produced for this purpose. 
The use of machine learning algorithms for image segmentation and classification 
is well established and has shown promising results in the case of identifying 
Type II and Type III solar radio bursts. Here we present the results of a 
convolutional neural network applied to dynamic spectra of NenuFAR solar 
observations. We highlight some initial success in segmenting radio bursts from 
the background spectra and outline the steps necessary for burst classification.
\end{abstract}

\section{Introduction}
\label{sec:intro}
Modern radio telescopes such as NenuFAR \citep[New Extension in Nançay Upgrading 
LOFAR;][]{zarka_lssnenufar_2012} generate data at rates of hundreds of gigabits 
per second, while next generation telescopes such as the Square Kilometre Array 
are expected to reach $\sim 1$ TB per second \cite{Scaife2020}. Solar radio 
bursts (SRBs) are bright, transient emissions and are classified in dynamic 
spectra by their drift in frequency. Many SRBs exhibit fine scale spectral 
features \citep[e.g.][]{Carley2015,Sharykin2018, Clarke2019} that require the 
highest spectral and temporal resolutions to be analysed. Like many solar 
phenomena, the occurrence of SRBs is not determined \textit{a priori} and as 
such, solar observations are likely to contain many hours of only background 
noise from the sky.
Due to data storage limitations, it is impractical to save all solar data 
recorded at the highest temporal and spectral resolution. It is also impractical 
to manually determine which observations should be archived at full temporal and 
spectral resolution and which should be rebinned or even deleted. 
Furthermore, large catalogues of SRBs exhibiting fine scale structure are lacking 
in the literature which presents a significant obstacle in understanding the 
statistical properties of these bursts. Thus, the development of a machine 
learning (ML) algorithm to automatically determine when SRBs occur is a vital tool
not only in the management of these observations, but also in filling in the gaps 
of our knowledge of the underlying physics of SRBs.

In this study we use a Convolutional Neural Network (CNN), a type of ML 
algorithm, to locate and identify SRBs in NenuFAR dynamic spectra. CNNs have 
previously been used in the context of SRBs and have seen success in  detecting 
and classifying type II and type III radio bursts \citep{Scully2021}.
UNET \citep{ronneberger2015} is a CNN that can classify an object on  a per pixel 
basis, this is known as image segmentation. UNET consists of a 
contracting/encoder path which downsamples the input to generate a number of 
feature, followed by an expansive/decoder path which upsamples these features to 
produce the model output \citep{ronneberger2015}.

Here we use UNET to perform segmentation on NenuFAR dynamic spectra and detect 
the exact time and frequency occurrences of SRBs. We use both Stokes I and Stokes 
V dynamic spectra as the input to the model. Section \ref{sec:method} describes 
how the dynamic spectra are prepared for use with UNET and how it was trained. 
The ML model's performance and qualitative results are shown in section 
\ref{sec:results}. In section \ref{sec:discussion} we discuss how our model's 
results can be used to qualify its success before finally summarising and 
outlining some future work in section \ref{sec:conclusions}.

\section{Method}
\label{sec:method}
In general, ML algorithms consist of neurons connected together in a layer. Each neuron has its own weight and bias and its output is determined by applying these to the output of neurons in the layer below. ML algorithms are trained by comparing the output of the model with a ground truth and calculating a loss function. The weights and biases of all the neurons in the model are updated in order to minimise the loss function and the procedure is repeated over a number of epochs until a minimum has been reached.

To avoid overfitting the model, the input data is split into a training set, used to learn underlying patterns, and a validation set, to test the predictions of the model. In supervised learning algorithms, both the training and validation sets need labelled annotations to compare as ground truths. For a segmentation algorithm such as UNET, the model determines a mask from a given input highlighting the location of a certain feature, in our case a SRB. 

Here we use UNET to generate masks of solar radio emission from NenuFAR observations. Currently , there exist no labelled ground truths to use and so they must be created before UNET can be trained. Below we describe how a labelled set of masks were created from the input dynamic spectra followed by the training procedure for UNET.

\subsection{Data Preparation}
In order to generate the ground truth masks for UNET, the onset and offset times of SRBs in NenuFAR dynamic spectra must be determined. The onset of bursts can be found by analysing the Poisson cumulative sum of a time series and has seen success in finding proton arrival times from SEP events \citep{PoissonCUSUM}. This technique is particularly useful for determining the onset of a single burst however, it does not perform well during Type III radio storms which contain hundreds of bursts over a number of hours.
In order to determine the onset and offset of multiple SRBs, we make use the cumulative sum slope \citep[CUSUM-slope;][]{tam_theoretical_2009}, 

$$
m_{t, \tau} = \frac{1}{\tau} \sum_{i=t+1}^{t+\tau}x_i,
$$

where $m_{t, \tau}$ is the CUSUM-slope at time $t$ for a time window of length $\tau$ and $x_i$ is the value of the time series at time $i$.

\begin{figure}[ht]
    \centering
    \includegraphics[width=0.75\columnwidth]{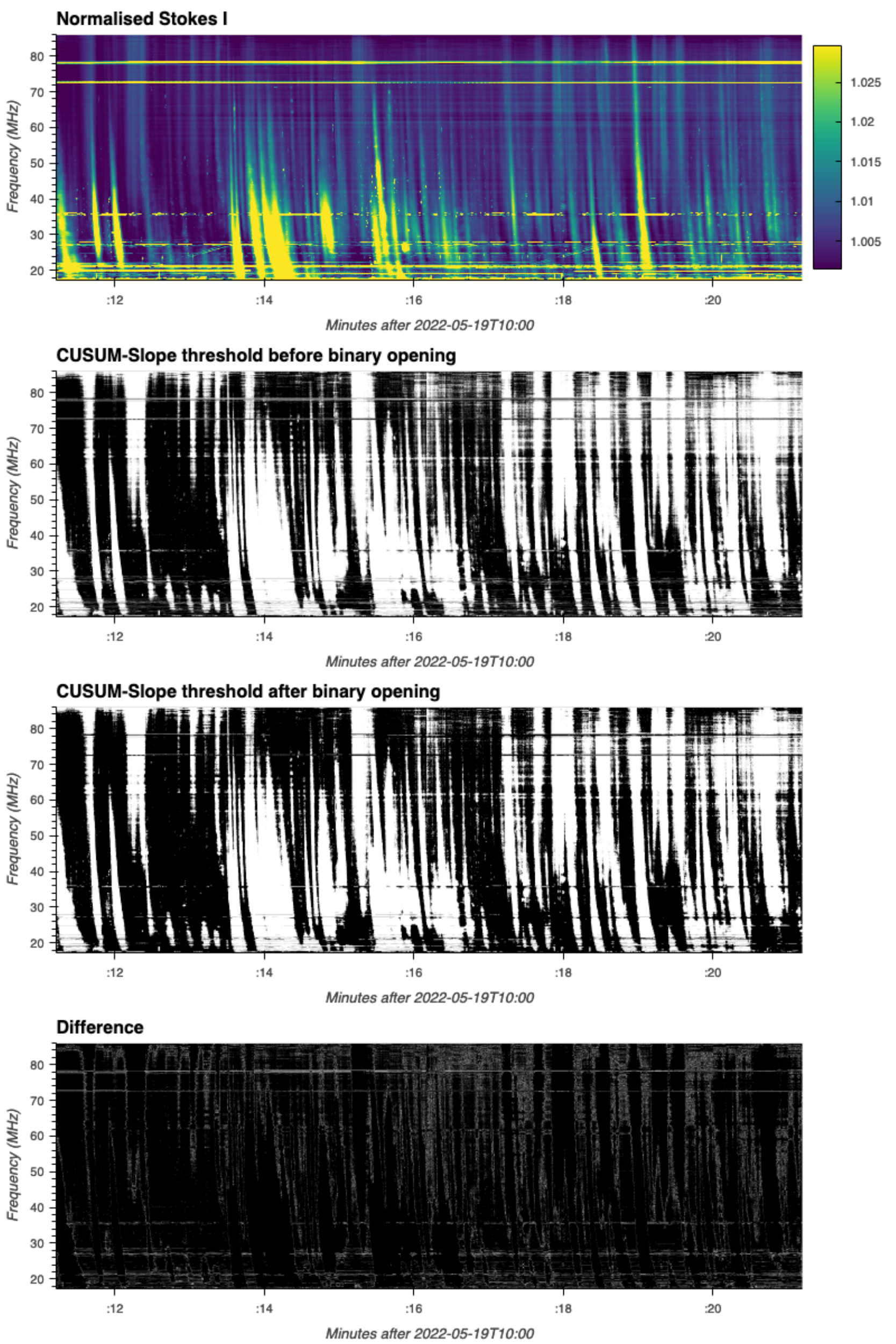}
    \caption{Demonstration of CUSUM-slope applied to Stokes I dynamic spectrum to generate a ground truth mask. The time axis starts at 2022-05-19T10:11:13. Top panel, Stokes I dynamic spectrum, the colour bar indicates normalised flux. The second panel from the top shows the mask determined by the CUSUM-slope threshold. The third panel is the result of a binary opening morphological operator applied to the mask from the middle panel. The bottom panel shows the difference in masks before and after the binary opening is applied.}
    \label{fig:masking}
\end{figure}

The CUSUM-slope is effectively a moving average of the data where the noise is zero-mean centred \citep{tam_theoretical_2009}. This allows a simple horizontal threshold to be used in order to detect SRBs. We generate a mask for both Stokes I and Stokes V dynamic spectra by classifying a CUSUM-slope signal greater than the 75\textsuperscript{th} percentile as an SRB. In order to clean the mask of spurious detections of radio frequency interference (RFI), a binary opening morphological operator is applied. We choose a horizontally oriented $1 \times 4$ pixel structuring element. This is large enough to remove small pixel sized noise but not so large that it removes real fine scale structure. The ratio of masked area before and after the binary opening step was 0.88 for the Stokes I spectrum shown in Figure \ref{fig:masking} and 0.20 for the corresponding Stokes V spectrum owing to the higher proportion of noise. Figure \ref{fig:masking} shows the procedure applied to a Stokes I dynamic spectrum on 2022-05-19T10:11:30.

Having created ground truth masks from a CUSUM-slope threshold to both Stokes I and Stokes V dynamic spectra, we have defined the two desired outputs from the UNET. The inputs for our UNET consist of three ``channels"; the dynamic spectrum for Stokes I, normalised to the background, the absolute value of the ratio of Stokes V to Stokes I, also normalised to the background, and the frequency of each channel in MHz.

\subsection{Training the model}
We train our UNET with observations from March to June of 2022 totalling approximately 50 hours. In order to create the inputs for UNET, we use dynamic spectra with a time resolution of $\sim 0.25$ seconds and a frequency resolution of $\sim 24.4$ kHz.
We split these into tiles of 256 time $\times$ 256 frequency steps which corresponds to a time/frequency range of $\sim 1\mbox{ min} \times 6$ MHz.
The tiles are shuffled and 20\% allocated as the validation set while the rest are used for the training set.
In order to increase the amount of training data, we apply a random shift to the starting time and frequency of the tiles in the training set.

A technique employed in training CNNs is to apply the weights of a CNN trained on a different but similar task to your CNN. This is known as transfer learning. Here we transfer the weights from a UNET trained on ImageNet \citep{deng_imagenet_2009}, a database of millions of labelled images, to the encoder path of our UNET. 

The loss function used during training was the sum of binary cross entropy, $L_{bce}$, and Jaccard loss, ${L_J}$,

$$
L = L_{\mbox{bce}} + L_{\mbox{J}},
$$

where 

$$L_{\mbox{bce}} = -y \log(y')-(1-y)\log(1-y'),$$ 

and

$$ L_{\mbox{J}} = 1 - \frac{y \cap y'}{y \cup y'} \,,$$

where $y$ is the ground truth and $y'$ is the prediction from the model.

We begin training the UNET with the frozen encoder weights from ImageNet for 100 epochs. Next, the encoder weights are unfrozen and the model is trained again with a lower learning rate for 50 epochs. In both training cycles we employ an early stopping regularisation, i.e. if the loss on the validation set does not improve for 5 epochs in a row, the training is stopped. The UNET model used here was implemented using the \texttt{segmentation\_models} Python library \citep{Yakubovskiy:2019}. The total size of the data used in training the model is 340 GB. The model was trained using an NVIDIA Tesla V100-PCIE-32GB and took 6.1 hours to train for 35 epochs. The code used to prepare and train the model are available at \url{https://gitlab.obspm.fr/pmurphy/nenufar_ml}.

\section{Results}
\label{sec:results}
\begin{figure}[ht]
    \centering
    \includegraphics[width=\columnwidth]{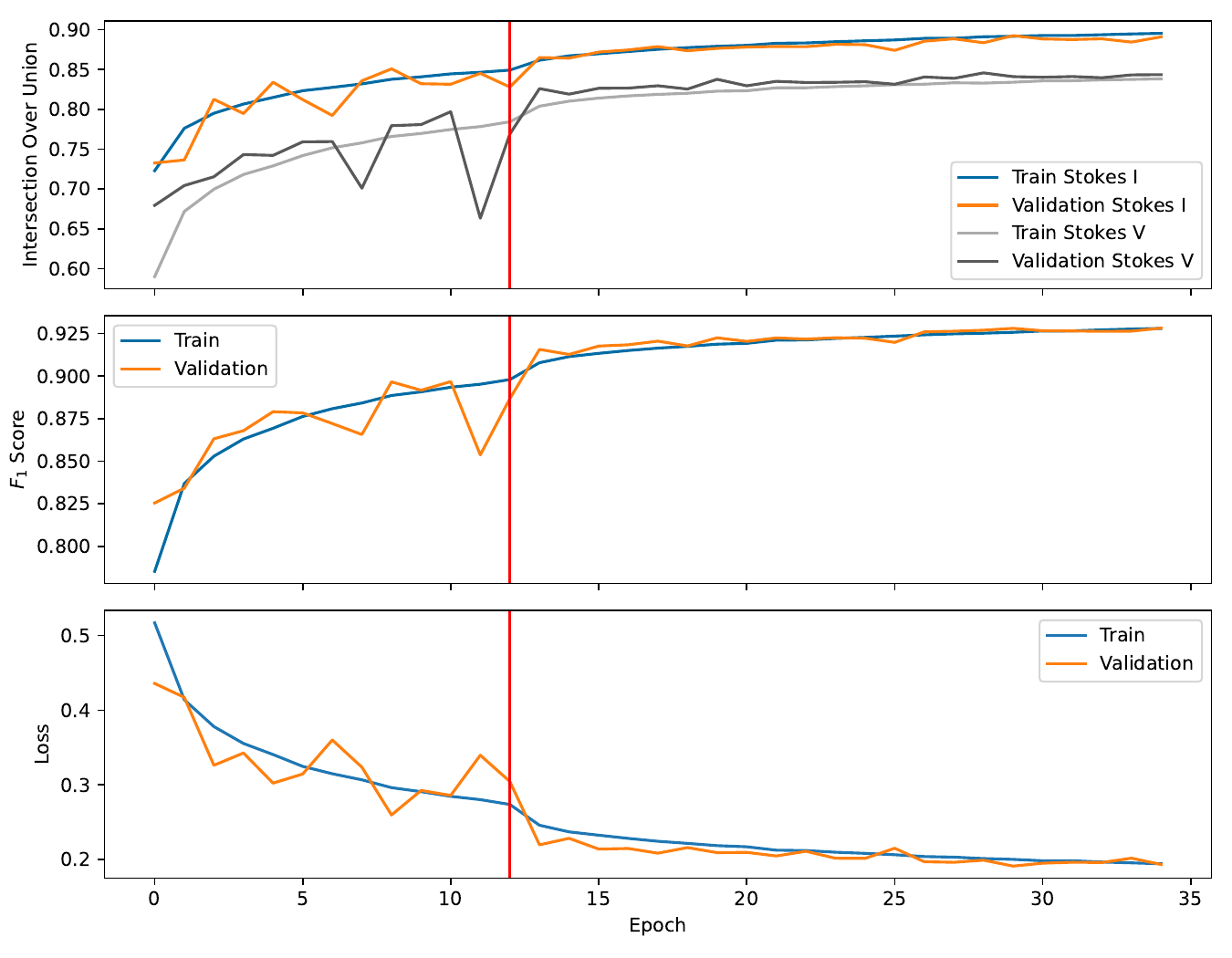}
    \caption{Model performance over training epochs. The top panel shows the intersection over union (IOU) for the training and validation set for both classes with increasing training epochs. The middle panel shows the $F_1$ score over training epochs for both classes and the bottom panel shows the loss function with respect to training epoch. The vertical red line at 12 epochs is the point where the encoder weights were unfrozen.}
    \label{fig:losscurve}
\end{figure}

While the ML algorithm is designed to minimise the loss function, a number of metrics can be used to evaluate the model's ability to make predictions. The most relevant in our case is known as the Intersection over Union (IOU) which is a metric to determine the similarity between the ground truth and prediction. As the name suggests, IOU is calculated by dividing the intersection of two sets by their union

$$
\mathrm{IOU} = \frac{y \cap y'}{y \cup y'} \,,
$$

where $y$ and $y'$ are the ground truth and model prediction as before. An IOU of 1 indicates perfect recreation whereas an IOU of 0 means there is no similarity between the ground truth and prediction.
The top panel of Figure \ref{fig:losscurve} shows the IOU calculated for both Stokes I and Stokes V masks separately. It is clear that both improve with each training epoch however, we note the IOU for Stokes V masks is not as high as for Stokes I and will discuss this further in section \ref{sec:discussion}.

A metric that is used as a proxy for the overall accuracy model is known as the $F_1$ score which we show in the middle panel of Figure \ref{fig:losscurve}. The $F_1$ score is the harmonic mean of a model's precision, accuracy of positive predictions, and recall, ratio of positive predictions correctly made. The $F_1$ is given as

$$
F_1 = 2 \frac{ \mathrm{precision} \cdot \mathrm{recall} }{ \mathrm{precision} + \mathrm{recall} } = \frac{ 2\mathrm{TP} }{ 2 \mathrm{TP} + \mathrm{FP} + \mathrm{FN} } \,,
$$

where 

$$
\mathrm{precision} = \frac{\mathrm{TP}}{\mathrm{TP} + \mathrm{FP}} \,,
$$

$$
\mbox{recall} = \frac{\mathrm{TP}}{\mathrm{TP} + \mathrm{FN}} \,,
$$

and TP stands for true positives, FP for false positives and FN for false negatives.

The bottom panel of Figure \ref{fig:losscurve} shows how the loss decreases with each epoch. The sharp decrease in loss at 12 epochs, marked with a red vertical line in all panels of Figure \ref{fig:losscurve}, is due to unfreezing the encoder weights.

\begin{figure}[ht]
    \centering
    \includegraphics[width=\columnwidth]{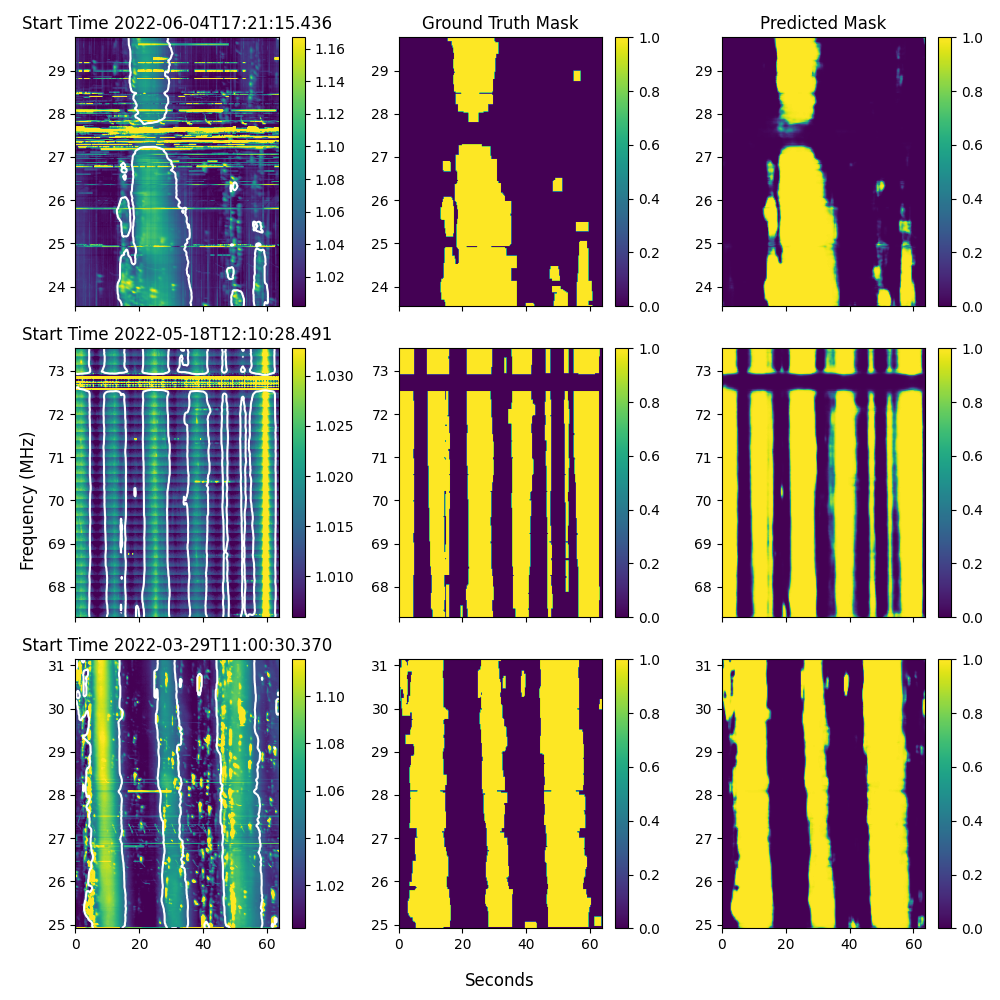}
    \caption{A qualitative comparison of the ML model predictions for Stokes I dynamic spectra. The left column shows the input dynamic spectrum, the middle column is the ground truth mask determined by the CUSUM-slope and the right column is the prediction from UNET. Each row shows an example at a different time and frequency range. White contours in the left column show where the predicted mask is above 80\%.}
    \label{fig:StokesI}
\end{figure}

\begin{figure}[ht]
    \centering
    \includegraphics[width=\textwidth]{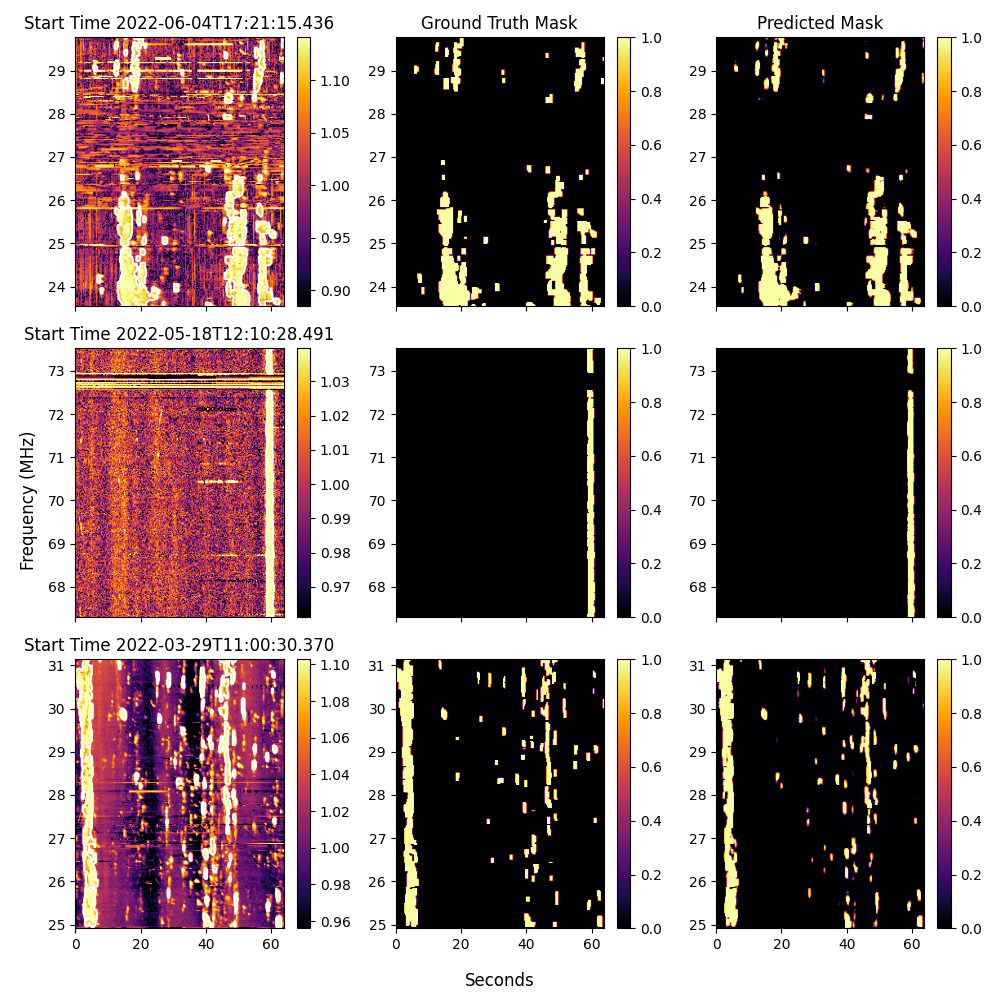}
    \caption{The same as Figure \ref{fig:StokesI} expect for Stokes V.}
    \label{fig:StokeV}
\end{figure}

A qualitative comparison between the input Stokes I dynamic spectrum, ground truth mask and predicted mask is shown in Figure \ref{fig:StokesI}. Here each row depicts the spectrum, ground truth and predicted mask for the same starting time and frequency. The white contours in the first column are where the predicted mask is greater than 0.8. The same comparison is made for Stokes V and is shown in Figure \ref{fig:StokeV}. The model predictions show an overall similarity with the ground truth masks and crucially, do not determine areas of strong RFI (e.g. 27-28 MHz) as SRBs.

\section{Discussion}
\label{sec:discussion}
Figure \ref{fig:losscurve} shows that after only 35 training epochs, our UNET has reached a minimum in the loss function and returned an $F_1$ score of greater than 90\%. However, there are a number of things that must be addressed before we can call this experiment a success. First, we note the different IOU scores for Stokes I and Stokes V masks. The difference between the two means that the model is not as effective at determining SRBs in Stokes V as in Stokes I. We hypothesise that this is due to fewer SRBs with strong Stokes V polarisation which leads to a class imbalance in the training and validation sets. For example, fundamental emission of Type III SRBs has stronger circular polarisation than harmonic emission \citep{Dulk1980}. Another reason for the lower IOU score in Stokes V masks is that Stokes V emission is commonly observed in fine scale features, e.g. Type IIIb striations \citep{morosan_exploring_2022} and fine scale features in noise storms \citep{mugundhan_spectropolarimetric_2018}. Thus, Stokes V emissions tend to be smaller features and a lower IOU score is indicative of the model struggling to recreate these small structures exactly.

A number of algorithms \citep[e.g.][]{zhang_rfi_2023, offringa_post-correlation_2010} make RFI removal a relatively easy task to perform. However, we decided not to perform RFI removal so that the model could learn to ignore RFI inherently. The top and middle panels of Figure \ref{fig:StokesI} show that despite bands of RFI at $\sim 27 - 28$ MHz and $\sim 72.5 - 73$ MHz, the model successfully ignores them when reproducing the segmentation mask. Removing the RFI before creating the ground truth masks could remove the need for the binary opening step and is worth exploring in the future.

It is also evident that the curves for the validation set exhibit more noise than the training set, particularly while the encoder weights are frozen. This suggests that tiles in the training set are not representative enough of those in the validation set. It is also possible that shifting causes some tiles in the training set to overlap with those in the validation set which would cause some batches to perform better than others. Applying a decay to the learning rate, such that the learning rate decreases with increasing epochs should reduce the noise in the validation loss.


Figures \ref{fig:losscurve} - \ref{fig:StokeV} show the quantitative and qualitative performance of the UNET employed in this study. We are confident that our UNET can successfully recreate the ground truth masks for a given dynamic spectrum.  However, it is important to note that this is not necessarily the same as saying that the UNET succeeds in detecting all SRBs in NenuFAR observations. In particular, care must be taken in the binary opening step so that the operation does not remove real fine scale structure along with pixel-sized noise. We believe our choice of structuring element described in section \ref{sec:method} is appropriate, although, it is apparent that smaller scale features in the middle columns of Figures \ref{fig:StokesI} and \ref{fig:StokeV} are ``boxy".
Thus we consider our results as a successful first step to creating a more complex algorithm. Nevertheless, the current predicted masks can be used to make some high level decisions on existing observations. For example, a ratio of burst to non-burst pixels in a given time/frequency range can be used to approximate what temporal and spectral resolution is necessary to retain relevant information. While further restraints e.g. the time range must contain both Stokes I and Stokes V SRBs, can inform the storage requirements on a \textit{per user} basis.

\section{Conclusions and Perspectives}
\label{sec:conclusions}
Using $\sim 50$ hours of NenuFAR solar observations, we trained a UNET to identify times and frequencies in dynamic spectra containing solar radio bursts. The trained model has an $F_1$ score of 93\% and recreates masks for SRBs in Stokes I and V with an IOU of 89\% and 84\% respectively. These results, along with the qualitative results shown in Figures \ref{fig:StokesI} and \ref{fig:StokeV}, highlight the success of our UNET in recreating the ground truth masks. 

We caution however, that this is only the first step in creating an ML algorithm to automatically determine whether data should be saved at its highest temporal and spatial resolutions. More detailed annotations of NenuFAR dynamic spectra would allow for the recognition of fine scale spectral/temporal structure and also to identify the five canonical solar radio burst types.

Given our success using a UNET with little customisation, we believe that a more sophisticated ML model could lead to the real-time detection of SRBs. With such a system, it would be possible for solar radio observations to ``piggyback" off of other observations and flag any SRBs that may be observed.
Furthermore, using a CNN to perform instance segmentation would produce masks for individual SRBs in each dynamic spectrum rather than regions of multiple SRBs. This would be essential in populating catalogues for the further study of the statistical properties of SRBs.

\section*{Acknowledgements}
This work has been supported by the Europlanet-2024 Research Infrastructure project, which received funding from the European Union's Horizon 2020 research and innovation programme under grant agreement No 871149 and a grant from Paris Observatory science council under the MINERVA project.

\newcommand{\newblock}{}
\bibliographystyle{mnras}
\bibliography{references.bib}

\end{document}